\documentclass[10pt,a4paper,final,journal]{IEEEtran}
\usepackage[utf8]{inputenc}
\usepackage[noadjust]{cite}
\usepackage{xcolor}
\usepackage{multicol}
\usepackage{amsmath}
\usepackage{amsfonts}
\usepackage{amssymb}
\usepackage{caption}
\usepackage{enumitem}
\usepackage[hidelinks]{hyperref}
\usepackage{graphicx}
\usepackage{bm}
\usepackage{geometry}
\usepackage{float}
\usepackage{wrapfig}
\usepackage{titling}
\usepackage{stfloats}
\usepackage{array}
\usepackage{xcolor}
\usepackage[normalem]{ulem}
\newcommand{\fig}{Fig.}
\newcommand\redsout{\bgroup\markoverwith{\textcolor{red}{\rule[0.5ex]{2pt}{0.4pt}}}\ULon}
\newcolumntype{P}[1]{>{\centering\arraybackslash}p{#1}}
\newcolumntype{M}[1]{>{\centering\arraybackslash}m{#1}}

\begin{document}
\title{A real time digital receiver for correlation measurements in atomic systems}

\author{\IEEEauthorblockN{V. Mugundhan\IEEEauthorrefmark{1},
Maheswar Swar\IEEEauthorrefmark{2}, Subhajit Bhar\IEEEauthorrefmark{3} and
Saptarishi Chaudhuri\IEEEauthorrefmark{4}}
\\\
\IEEEauthorblockA{Raman Research Institute,\\
C.V. Raman Avenue, Sadashivanagar, \\
Bangalore - 560080, India\\\
\IEEEauthorrefmark{1}mugundhan@rri.res.in,
\IEEEauthorrefmark{2}mswar@rri.res.in,
\IEEEauthorrefmark{3}subhajit@rri.res.in,
\IEEEauthorrefmark{4}srishic@rri.res.in}}

\date{}

\twocolumn[
  \begin{@twocolumnfalse}
    \maketitle
    
  \end{@twocolumnfalse}
]
\begin{abstract}
     We present the development and characterization of a generic, reconfigurable, low-cost ($<$ 350 USD) software-defined digital receiver system (DRS) for temporal correlation measurements in atomic spin ensembles. We demonstrate the use of the DRS as a component of a high
resolution magnetometer. Digital receiver based fast Fourier transform spectrometers (FFTS) are generally superior in performance in terms of signal-to-noise ratio (SNR) compared to traditional swept-frequency spectrum analyzers (SFSA). In applications where the signals being analyzed are very narrow band in frequency domain, recording them at high speeds over a reduced bandwidth provides flexibility to study them for longer periods. We have built the DRS on the STEMLab 125-14 FPGA platform and it has two different modes of operation:  FFT Spectrometer  and real time raw voltage recording mode. We evaluate its performance by using it in atomic spin noise spectroscopy (SNS). We demonstrate that the SNR is improved by more than one order of magnitude with the FFTS as compared to that of the commercial SFSA.  We  also highlight that with this DRS operating in the triggered data acquisition mode one can achieve spin noise (SN) signal with high SNR in a recording time window as low as 100 msec. We make use of this feature to perform time resolved high-resolution magnetometry. While the receiver was initially developed for SNS experiments, it can be easily used for other atomic, molecular and optical (AMO) physics experiments as well.
\end{abstract}


%

\section{\label{s1}Introduction}

Digital receivers are a class of electronic systems where operations like amplification, filtering, integration etc. are performed as a series of mathematical operation on embedded components like FPGAs, Microprocessors or GPUs. Compared to their analog counterparts, digital receivers are immune to variations in gain and temperature. However, digital systems have quantization noise, sampling rate and phase noise which can be
minimized by choosing high bit-width ADCs and low drift clock sources. These characteristics make digital receivers an attractive option in applications where precision measurements are required \cite{wepman1996,jamin2014}. AMO experiments are one such example.   

\par
Digital receivers are being used in AMO experiments for process control \cite{luda2019} (e.g. temperature, current and wavelength control in lasers), synchronous detection \cite{stimpson2019} and in compact magnetic resonance spectroscopy systems \cite{takeda2007}. However, synchronous detection may not be feasible in many AMO experiments including spin noise spectroscopy (SNS) \cite{MSwar2018, zapasskii2013, Crooker2004}. The typical strength of the raw SNS signal is $< 100nV/{\sqrt{Hz}}$ which is far less compared to the previous studies\cite{luda2019,stimpson2019,takeda2007}. Therefore, the SNS signal has to be recorded continuously (or on trigger) to improve the SNR.
 
\par
In this paper, we discuss the development and utilization of a versatile digital receiver to measure the spin noise (SN) in atomic vapor systems. We have performed a comparative study with the results of our previous work in \cite{MSwar2018}. We also demonstrate its utility in real-time
precision magnetometry. This digital receiver will be used as a component of a novel, miniaturized magnetometer based on SNS techniques. 
\par
The paper is organized as follows: We introduce spin noise spectroscopy and the digital receiver system developed for its measurement in section
\ref{s2}. The firmware architectures of the former is described in section \ref{s3}. The methods adopted to mitigate the effects of electro-magnetic interference (EMI) in the measurements are described in section \ref{s4}. In section \ref{s5}, we present the SNS data obtained by using our DRS as well as a comparison with the results obtained from SFSA. Further, we demonstrate triggered data acquisition and time resolved magnetometry using our DRS. We conclude in section \ref{s6} after a brief discussion on further applications and future scope of the developed receiver.

\section{Digital Receivers for Spin noise spectroscopy}\label{s2}
In this section, we describe the spin noise spectroscopy technique as well as detection schemes by introducing our digital receiver.
\subsection{Spectroscopy Technique}
Study of SN of an atomic ensemble has varied applications, ranging from precision magnetometry, non-perturbative optical detection to metrology and quantum sensing. The fluctuation in the spin population of an atomic system in time leads to temporal changes in the refractive index of the medium. A linearly polarized and far detuned probe laser beam  can detect this temporal variations of the refractive index in its polarization angle. We use a polarimetric detection system where the polarization fluctuation of the probe laser beam is detected in a balanced photodetector \cite{MswarBook}. The power spectrum of the polarization fluctuation (second order correlation function - g$^{(2)}$) gives the information about the spectral properties of the atomic spin ensemble. In a typical experimental conditions we apply a uniform magnetic field perpendicular to the propagation direction of the probe beam. The atomic spins precess around the magnetic field with a Larmor frequency which is proportional to the strength of the magnetic field.  Therefore, the peak of the spectrum appears at the Larmor frequency. Also, the width of the signal is inversely proportional to the transverse spin relaxation time ($\sim$100 kHz in our experimental conditions). 

\begin{figure*}[!ht]
    \centering
    \includegraphics[trim={3.5cm 6cm 2.5cm 2cm},clip,scale=0.55,keepaspectratio]{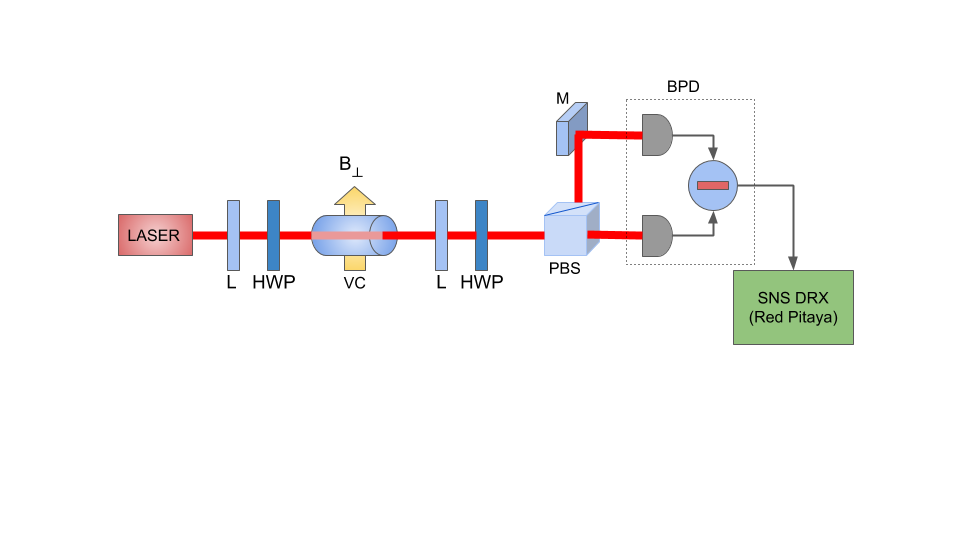}
    \caption{A typical spin noise spectroscopy (SNS) set-up of rubidium (Rb) atomic vapor. L- plano-convex lens, VC- vapor cell (contains Rb atomic vapor), M- dielectric mirror, HWP- half wave plate, PBS- polarizing beam-splitter, BPD- balanced photodetector, $B_{\bot}$- orthogonal magnetic field.}
    
    \label{F1}
\end{figure*}

Our SNS experimental set-up is shown in \fig\ref{F1}. A probe laser beam which is detuned by $\approx$ 10 GHz from the strongest optical transition in neutral rubidium (Rb) atoms, is used to probe the spin fluctuations of the atoms in the vapor cell heated to temperatures ranging between 350 K and 400 K. The spin fluctuations causes polarization fluctuations of the probe laser. The polarimetric detection scheme employing a half wave plate (HWP), polarizing beam splitter (PBS), and a balanced photo-detector (BPD) is capable of measuring this polarization fluctuation. A uniform magnetic field ($B_{\bot}$), produced using a pair of magnetic coils in Helmholtz configuration, is applied on the atomic vapor which is perpendicular to the propagation direction of the probe laser beam. The output signal of the BPD is recorded using the digital receiver described in this article and reported in section \ref{s5}.

Previously we used a commercially available SFSA\footnote{https://www.keysight.com/in/en/assets/7018-01953/data-sheets/5989-9815.pdf} to detect the SN signal from the balanced photo-detector. It has a superheterodyne stage whose mixer provided a swept local oscillator (LO) signal, such that the radio-frequency (RF) signal is translated to a fixed intermediate frequency (IF). Since the LO has to be swept across a range of frequencies, the sweep time increases with frequency span. This reduces the dwell time at each frequency resulting in a decreased sensitivity. From the experimental point of view, this leads to poor SNR. Moreover, if the signal is expected to show variations smaller than the sweep-time, it cannot be detected. SFSAs also have a low frequency cut-off (in our case 100 kHz) below which the output amplitude and frequency measurements are not possible. We have ongoing experiment to perform SNS on laser cooled atoms to investigate spin dynamics in quantum regime, where a triggered data recording of a SN signal over a short duration ($\sim$ few ms) is required. In this case a SFSA can not be used.

There have been recent experiments reporting measurements of SNS in quantum dots \cite{ crooker2010} as well as in atomic vapors \cite{lucivero2016}, where the use of non-reconfigurable and somewhat expensive digital receivers are reported. However, the digital receiver described in this paper, allows us to overcome the aforementioned limitations of SFSA, with certain trade-offs. 

\subsection{Digital Receivers for SNS}
We developed a digital receiver capable of operating in two modes, as listed below.

\begin{enumerate}[label=\Roman*]

    \item A fast Fourier transform spectrometer (FFTS) to probe the entire frequency range of interest,

    \item A real time data recorder (RTDR) with trigger capabilities.
    
\end{enumerate}

Both the aforementioned modes are implemented on the STEMLab 125-14\footnote{https://www.redpitaya.com/f130/STEMlab-board} development board. This board is selected for our application because it has two 14 bit analog-to-digital converter (ADC) channels, with each channel providing a dynamic range better than 80 dB. It has an analog bandwidth of 62.5 MHz, and is DC coupled. The heart of the board is a Xilinx Zynq 7010 System on Chip (SoC), with integrated programmable logic (PL) cells and ARM microprocessor based processing system (PS). The signal processing algorithms in this work are implemented on the PL side, while user control and data transmission programs are implemented on the PS side. 

\section{Firmware Description}\label{s3}
In this section, we describe the firmware architecture of the two operational modes of the DRS.
\subsection{Fast Fourier Transform Spectrometers (FFTS)}
Fourier transform is used to find the spectral content of a time domain signal \cite{bracewell1986}. Fast Fourier transform (FFT) is an algorithm which reduces the complexity involved in calculating the Fourier transform from a $O(n^2)$ to $O(n\log{n})$ problem by using the periodicity and the symmetry properties of the former \cite{cooley1969,duhamel1990}. This in general reduces the number of operations required to obtain the spectrum and results in resource savings when implemented in embedded devices e.g. FPGA or in Microprocessors.

The SNR is proportional to $\sqrt{\beta\tau}$, where $\beta$ is the bandwidth, and $\tau$ is the integration time \cite{tiuri1964}. For a conventional spectrum analyzer, there are two time scales involved: $t_s$, the sweep time, and $t_d$, the dead time. So, if a sweep contains $N_s$ points, the amount of time required for obtaining the power at each frequency becomes ${t_s}/{N_s}$. In cases where the data is to be acquired using interfaces such as GPIB, USB, or ethernet, $t_d$ includes the time taken for the spectrum analyzer to transfer the data to the DAQ system, during which time no new acquisition can occur.

In case of an FFTS, the estimation of the power spectrum involves complex weighting and summation of all time domain samples of the burst used to perform the FFT. If a streaming algorithm is used, data acquisition, performing FFT and data transfer can happen simultaneously resulting in zero dead time. Thus for a single spectrum, with the same spectral resolution, an FFTS ideally provides $\sqrt{N_s}$ improvement in the SNR as compared to SFSA\cite{mugundhan2018}.

\begin{figure*}[!t]
    \centering
    \includegraphics[width=\textwidth,scale=0.75,keepaspectratio]{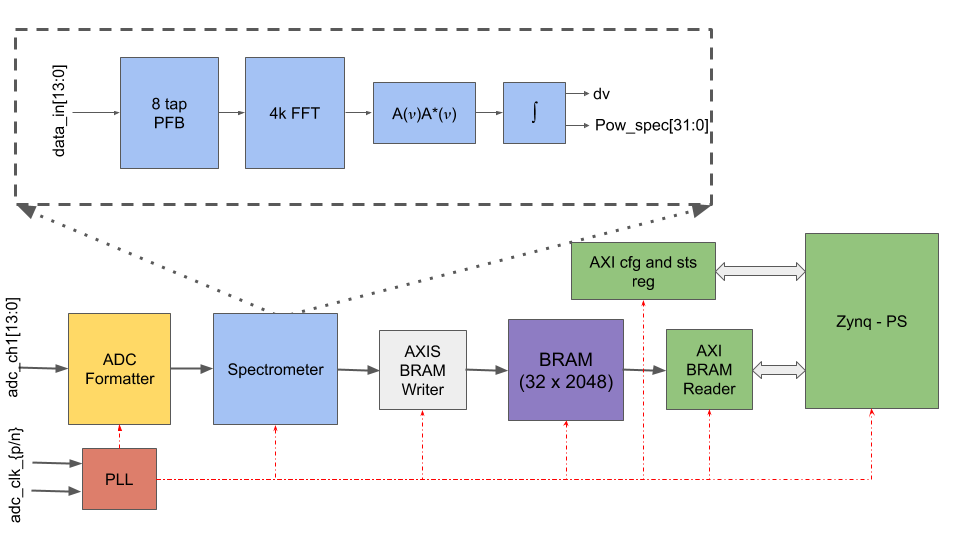}
    \caption{Top level block diagram of the FFTS. PS and related interfaces are shown as green blocks. The red dashed lines indicate the flow of the master clock at 125 MHz, which is derived from the ADC data clock. The data and clock inputs from the ADC to the FPGA are Low Voltage
Complementary Metal Oxide Semiconductor (LVCMOS) signals.}
    \label{F2}
\end{figure*}

The base firmware version of the Stem-Lab 125-14 provides a burst mode version of the FFTS. However, we required capability to perform this operation in streaming mode, and average the spectra on the FPGA itself. This allows us to keep data rates below 30 MBps, beyond which loss-less data transfer via ethernet becomes difficult, due to bottlenecks in communication between the PL and the PS sides of the SoC. The block diagram of our FFTS implementation is shown in \fig \ref{F2}.

The analog signal is sampled by the on-board ADC at 125 MHz. The digitized data is captured synchronously on the FPGA, converted to 2's complement format and passed on to the spectrometer block.

The spectrometer block is implemented using Simulink System Generator. The signed data obtained so far is polyphase filtered using an 8 tap FIR filter\cite{vaidyanathan1990}. The output of the filterbank is an 18 bit fixed-point number, with a binary point at the 17th bit. This is input to the biplex FFT block IP core \cite{emerson1976}, available from the CASPER signal processing library \cite{hickish2016,CASPERweb}. A 4096 points FFT is performed resulting in a spectral resolution of $\approx$ 30.5 kHz. To avoid overflow during subsequent stages of the FFT, the output of each stage is scaled by a factor 2. As the data to the FFT block is real-valued, the power spectrum is estimated by computing the magnitude of the positive half of the spectrum. After the integration of a programmable number of spectra, the \texttt{dv} signal is asserted high and the power spectrum is presented to the subsequent blocks for transmission to the DAQ. For example, if 1000 such spectra are summed, the resulting integration time is $\approx$ 32 ms. 

The output spectra is recast as an Advanced eXtensible Interface (AXI) stream and is written to a Block RAM (BRAM) through the \texttt{AXIS BRAM WRITER} IP block \cite{paveldemin}. Once the spectrum is written into the BRAM, a \texttt{finished} signal indicating this is asserted and posted to the \texttt{sts} register. 

The FPGA present on-board provides access only to the PS Ethernet, therefore the transfer of data to the DAQ is mediated by executing a C code on the PS. The \texttt{sts}, \texttt{cfg} and \texttt{AXI BRAM READER} provide memory mapped access to the PL. \texttt{cfg} register is used to provide a master reset and configuration information to the PL. \texttt{sts} register provides information on the BRAM address pointer and holds the state of the \texttt{finished} signal. On the assertion of the \texttt{finished} signal, the PS starts reading the contents of the BRAM. The BRAM data is packetized as \texttt{UDP} packets with packet and spectrum count information and is transmitted to the DAQ using linux socket functions.

We characterize  the developed spectrometer using continuous wave (CW) and SN signals. For the CW tests, signals at different frequencies and different powers are fed to the system and recorded. These tests are carried out to estimate the SNR of the spectrometer at different frequencies and to identify the linear regime. The power measured by the FFTS is found to linearly vary with the input power. 



\subsection{Real Time Data Recorder (RTDR)}

\begin{table*}[tp]
 \begin{center}
  \captionsetup{justification=raggedright}
   \caption{The different options available in the RTDR firmware.}\label{T1}
   \begin{tabular}{|l|l|}
  \hline
    \textbf{Option} & \textbf{Description} \\\hline
    baseband & Records signal in dc\,-\,625 kHz base-band; LO is disabled \\
    IF & Records signal in an IF band spanning from $f_{lo}$ to $f_{lo}+625$ kHz \\
        triggered & Records a signal burst for a programmable predefined time on rising edge of trigger pulse \\
    continuous & Records data continuously\\\hline
  \end{tabular}
   \end{center}
\end{table*}

The FFTS discussed thus far, provides a simple and compact measurement option for SNS. In scenarios where the SN signal is short lived in time, a real time triggered data acquisition and processing protocol is required.

\begin{figure*}[ht]
    \includegraphics[width=\textwidth,scale=0.75,keepaspectratio]{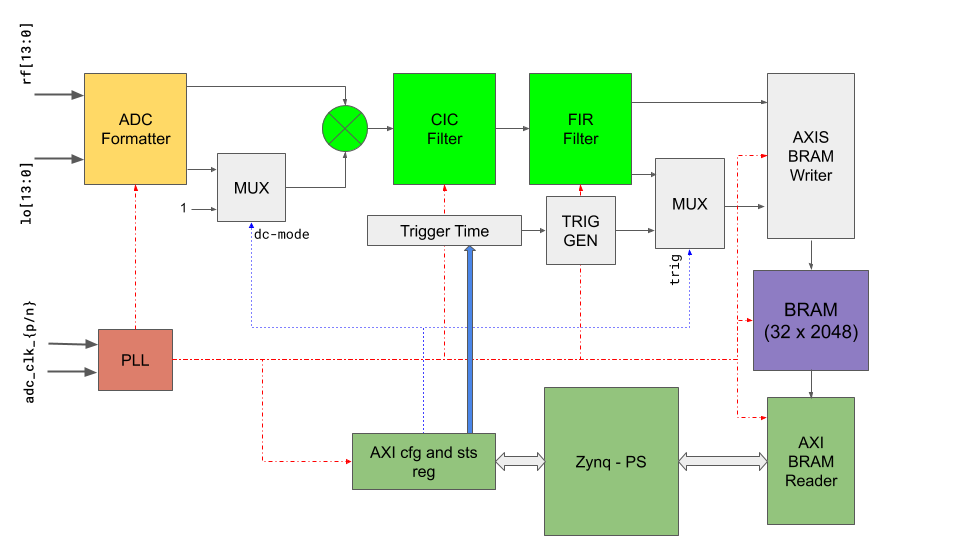}
    \caption{Firmware description of the triggered raw voltage recorder. The dotted blue lines represent the flow of control signals and status signals to and from the memory mapped AXI registers. The green blocks represent signal processing elements of the design.}
    \label{F5}
\end{figure*}

We developed a low bandwidth, raw voltage recorder. In \fig  \ref{F5}, we show a top-level block description of the same. Table \ref{T1} outlines the various options available in the data recorder. Here, we use both the input channels, where one is from the BPD and the other is a signal generator (BK PRECISION Model no. 4040B), which is used as the local oscillator (LO). As in the FFTS firmware, the signals from the ADC are converted to 2's complement and multiplied to obtain an intermediate frequency (IF) signal. The IF signal is usually close to dc and tracking the variations in the former using the master clock operating at 125 MHz would result in sub-optimal usage of resources. Hence, we use a cascaded integrated comb (CIC) filter for down-sampling the signal\cite{hogenauer1981,lyons2005}. When the factor by which the signal has to be down-sampled is large, CIC filters, used as a front end for FIR filters result in decreased number of filter taps required for anti-aliasing.

The response of the CIC filter is not uniform within the band of interest. To compensate for this, a FIR filter with a complementary response is cascaded to it. The FIR filter is implemented as a half-band filter, resulting in an output data rate that is half of that of the input. A 256 tap Kaiser window, with $\beta=10$, is used to shape the filter pass band. 
When both the filters are cascaded, the resulting combination has an uniform amplitude response over the band of operation and an out of band rejection $\geq$ 90 dB. 

The data from the multiplier as it enters the CIC filter is at 125 MSPS. This is decimated by 50, resulting in a data rate of 2.5 MSPS at the output of the CIC filter and an aggregate data rate of 1.25 MSPS after the FIR filter, resulting in a base-band data of 625 kHz.

We use four counters in the firmware for timekeeping purposes: a trigger counter (TC), a free running counter (FRC) at 125 MHz rate, a over-
flow counter (OC) and a packet counter (PC). The TC keeps track of the triggers received by the RTDR, while the FRC keeps track of the time
since the power on. The 32 bit OC counts the number of FRC overflows. These counters allow us to obtain the time between two trigger events
and also its occurance instances since the start of acquisition. The PC helps us to ensure no data was missed during packet framing and transmission.

\begin{figure*}[ht]
  \begin{center}
      \includegraphics[width=\textwidth]{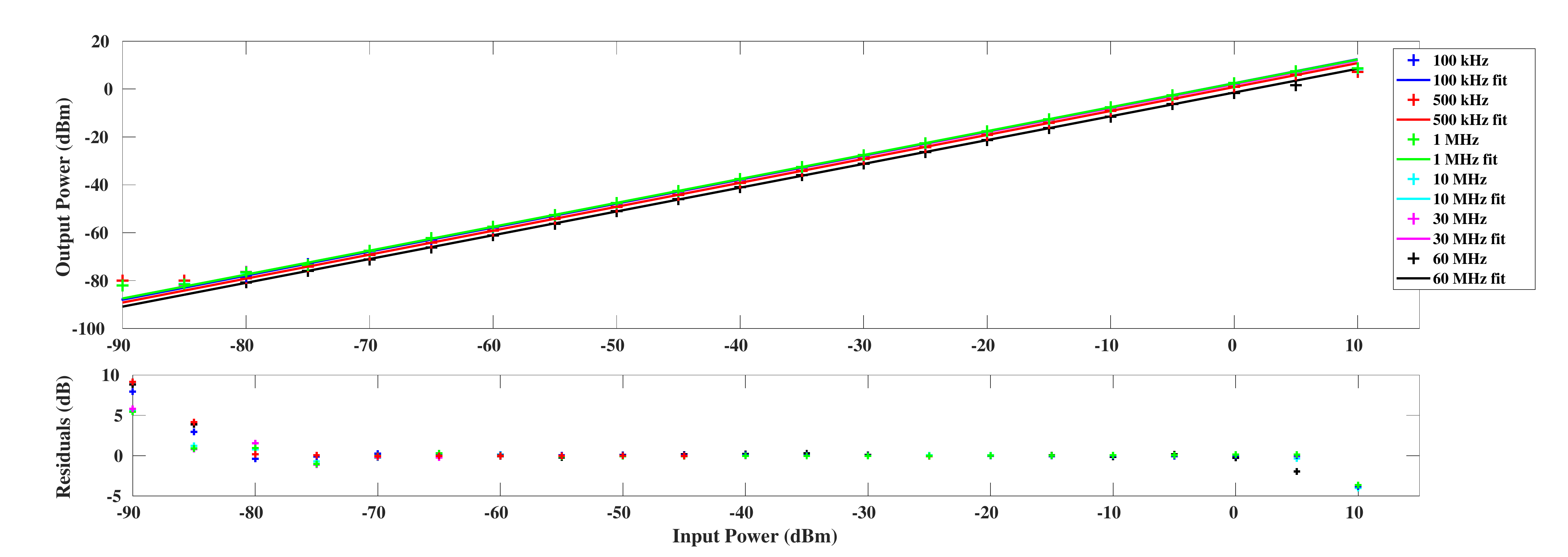}
  \caption{Characterization of the DRS (in both RTDR and FFTS modes) with signals of various power fed across the frequency range of operation. The '+' markers indicate the data and the solid lines represent the first order polynomial fits. The bottom plot shows the residuals. The black dashed lines encompass the linear range of the DRS.}
  \label{F13}
  \end{center}
\end{figure*}

\par
The data, along with the counter values, is written to a BRAM and read out by the PS and transmitted using ethernet. 

CW signals at various frequencies within the 0-62.5 MHz band were injected at different power levels from -90 to 10 dBm. As shown in Fig.\ref{F13}, the output power scales linearly with the input power, irrespective of the frequency.

\section{Electromagnetic Interference (EMI) and its mitigation}\label{s4}
\par
The ambient EMI can hinder the detection of weak signals. Some common sources of unavoidable EMI are 50/60 Hz AC lines, switching regulators and LO harmonics. Strong EMI affects the dynamic range of the receiver system at frequencies $<$ 1 MHz. EMI can be mitigated during pre- and post-processing stages. We describe the techniques adopted, during these stages of EMI mitigation, in this section.

\par
During our measurements, we found that the strong 50 Hz AC signal and its harmonics were getting coupled into the system through the AC adapter of the FPGA evaluation board. To overcome this, we replaced the power adapter with a commercially available battery-bank of similar specifications. 

\begin{figure}
  \centering
  \includegraphics[trim={0.5cm 0cm 0.5cm 0.5cm},width=\linewidth]{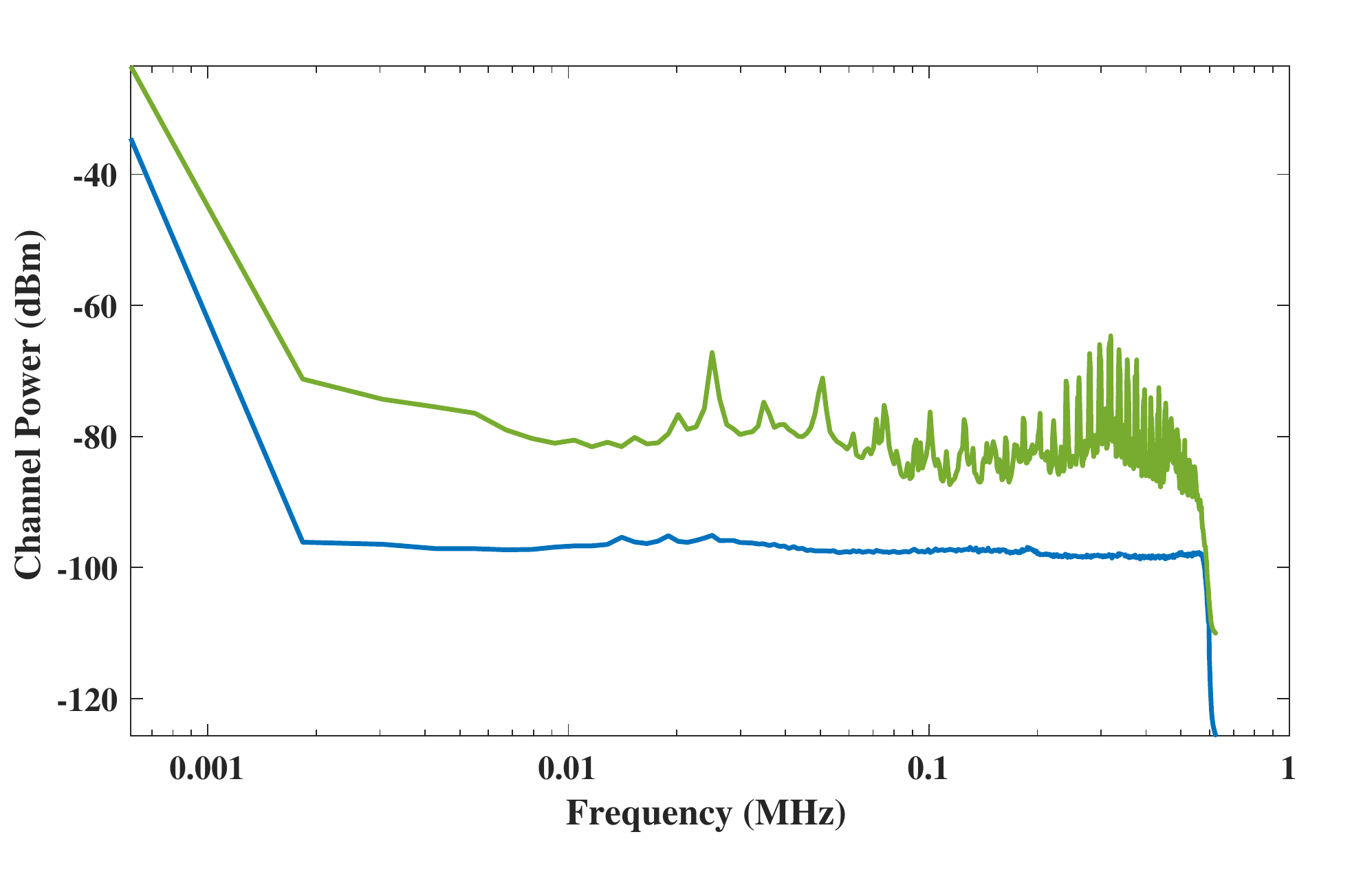}
  \caption{The effect of EMI on the measured spectrum (green trace) and after its mitigation (blue trace). The effect of EMI mitigation is clearly visible through the absence of spikes and a reduction in the noise floor
by $\approx$ 20 dB in the blue trace.}
  \label{F12}
\end{figure}

A second prominent source of EMI were the switching circuits associated with the magnetic coil current driver. This gave rise to strong peaks in the low kHz frequencies. The SNS setup is located in a laboratory environment with multiple sources of EMI. Therefore, we custom designed a mild-steel enclosure for the DRS. In \fig\ref{F12} we show the significant mitigation of EMI by $\approx$ 20 dB after adopting the aforementioned schemes. We performed further processing on the archived data to excise low-level narrow-band and impulsive broad-band EMI \cite{fridman2008},\cite{ford2014}. For EMI that was stationary in frequency, we used a combination of background and median subtraction. Impulsive, broadband EMI, when present, was clipped from each channel when its value exceeded the 3$\sigma$ threshold.

\section{Results and Discussion}\label{s5}
Our experimental set-up is shown schematically in \fig\ref{F1}. A uniform transverse magnetic field ($B_{\bot}$) was produced by the current flowing through a pair of magnetic coils in Helmholtz configuration. The signal from the BPD\footnote{The BPD used was a Newport 1807 model having a cut-off frequency of $\approx$ 80 MHz.} was recorded either with our digital receiver system (DRS) or a commercial spectrum analyzer (SFSA) for the purpose of performance comparison.
\par
\begin{figure}[!ht]
    \centering
    \includegraphics[trim={0.5cm 0.25cm 0.5cm 0.5cm},width=\linewidth]{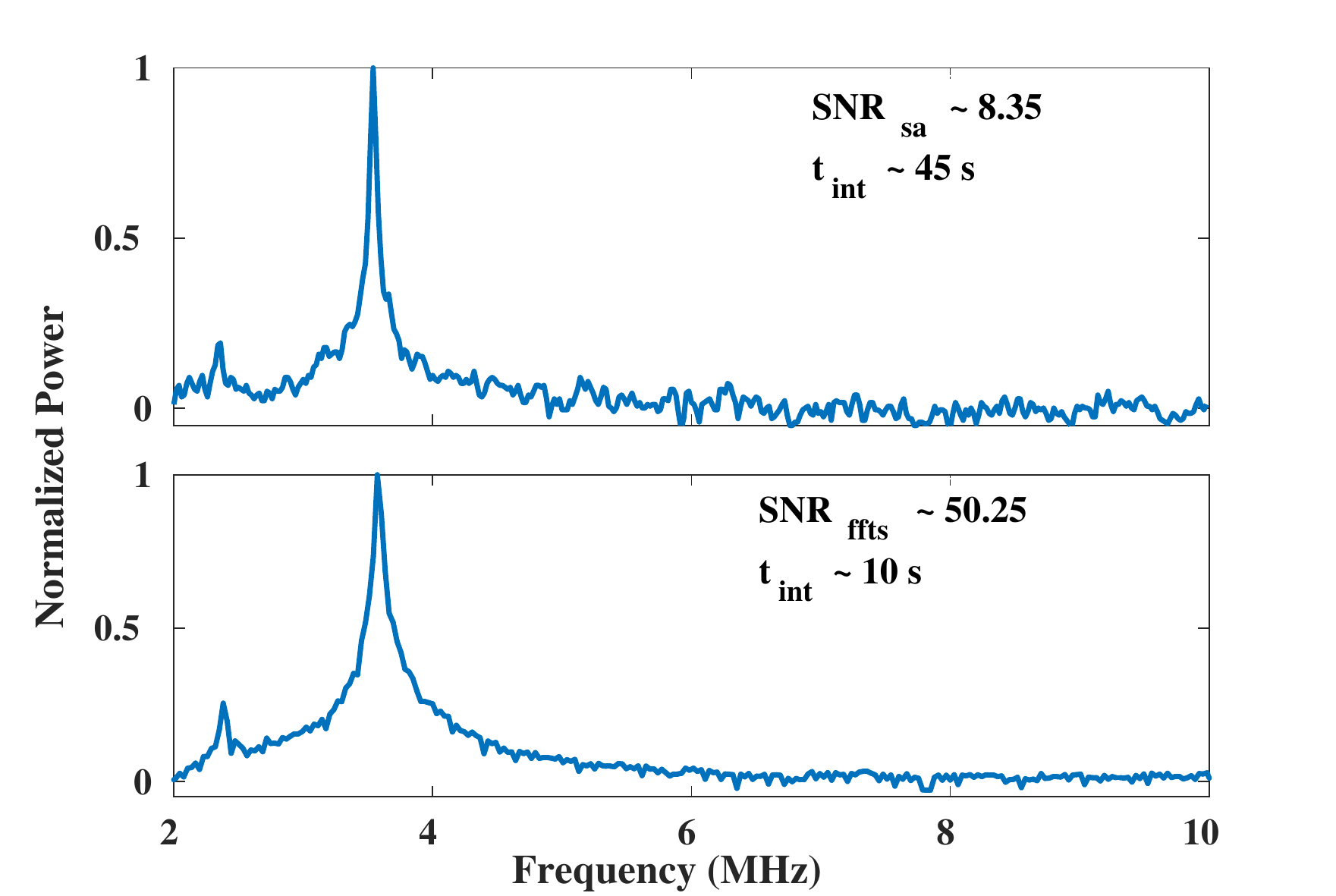}
    \caption{Spin noise (SN) spectrum acquired from SFSA (top panel) and FFTS (bottom panel). Note that the SNR is $\approx$ 50 for the FFTS for an integration time 5x lower than SFSA. These two spectra were recorded under the same experimental conditions, at $B_{\bot} \approx 5.1$ G.}
    \label{F8}
\end{figure}

In \fig\ref{F8}, we show the SN spectrum of Rb atomic vapor with $B_{\bot}$ $\approx$ 5.1 gauss obtained using SFSA (top panel) and the FFTS (bottom panel). These sets of data were recorded under similar experimental conditions. In each of the panels, we observe two peaks appearing at $\approx$ 2.4 MHz and $\approx$ 3.6 MHz corresponding to SN signal due to $^{85}$Rb and $^{87}$Rb, respectively. The RBW of the SFSA was the same as the channel width of the FFTS which is 30 kHz. The integration time for the SFSA to obtain the data presented in \fig\ref{F8} (top panel) was 45 seconds whereas the same for the FFTS presented in \fig\ref{F8} (bottom panel) was 10 seconds. In both the cases the spectrum was background subtracted and normalized to it's peak value. The SNR of the SN signal is defined as the ratio of the strength of the strongest signal to the rms value of the background signal. The background signal is obtained by recording the SN signal at zero magnetic field.  
Comparing these two spectra, we note that the SNR for the DRS is more than an order of magnitude better than that for the SFSA for the same integration time. This improvement in the SNR along with the fact that our DRS is light weight, portable, low-cost, low-power ($<$ 10 Watts) as compared to a commercial SFSA, makes it preferable for both laboratory and field measurements.   

\begin{figure}
    \centering
    \includegraphics[trim={0.5cm 0.25cm 0.5cm 0.5cm},width=\linewidth]{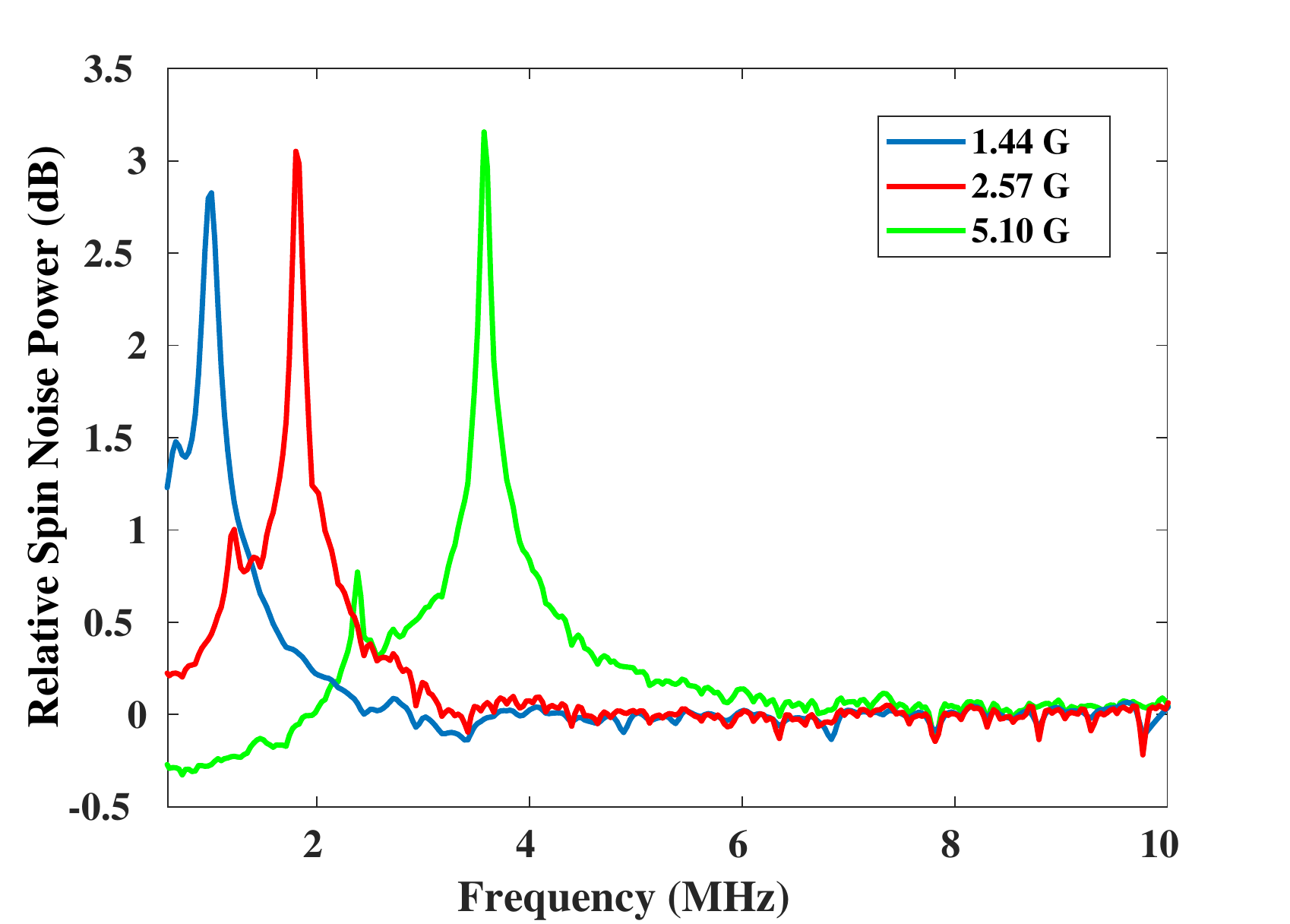}
    \caption{Background subtracted spin noise (SN) spectra measured for different magnetic field values. The error in the measurements are $\approx$ 0.02 G, which is limited by the spectral resolution $\approx$ 30 kHz of the FFTS.} 
    \label{F7}
\end{figure}

Since the SN spectrum peak position is the Larmour frequency $\nu_L (=g_F \mu_B B_{\bot} /h $, where $g_F$ is the g-factor of the hyperfine levels, $\mu_B$ is the Bohr magneton, and $h$ is the Plank's constant$)$, by precisely measuring the peak position of the spectrum we can estimate the strength of $B_{\bot}$. Therefore, this measurement technique can be used as a precision magnetometry tool. Since the developed DRS described in this article is easily field deployable, we would highlight the applicability of this device to construct a robust miniaturized magnetometer. As an example, the SN spectrum recorded at various magnetic field strengths are shown \fig\ref{F7}. By fitting a Lorentzian to the individual spectrum, we can estimate the Larmour frequency and in turn the magnetic field. 

\begin{figure}[!ht]
    \centering
    \includegraphics[trim={0.5cm 0.5cm 1cm 0.5cm},clip,width=\linewidth]{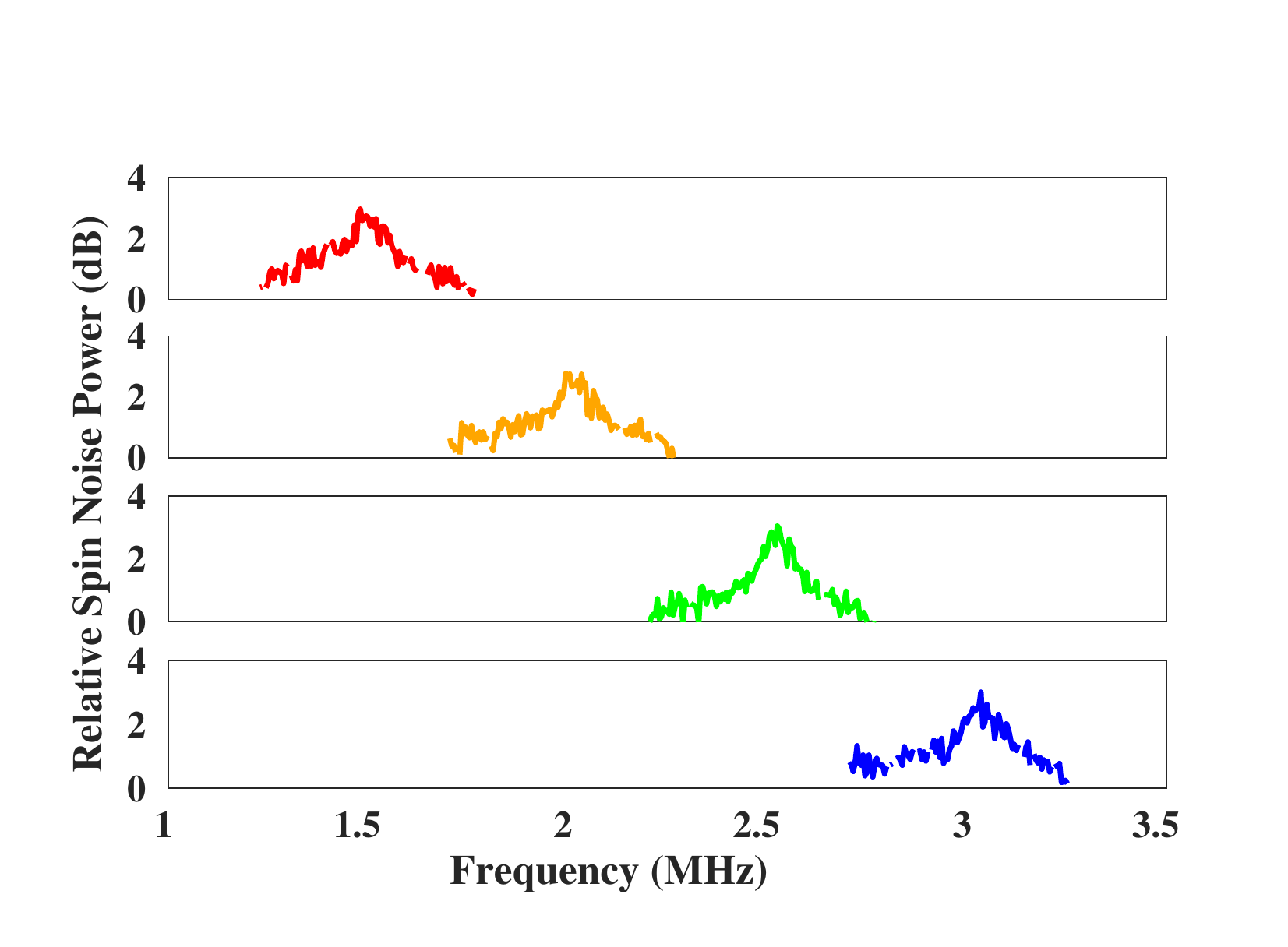}
    \caption{Spin noise (SN) spectrum at various magnetic field strengths from the RTDR. These series of spectra has an SNR $\approx$ 5, and the peak positions can be determined within an accuracy of $\approx$ 5$\%$.}
    \label{F8_realtime}
\end{figure}

Another advantage of using this DRS is that the device is reconfigurable which enables triggered real time measurements of the SN spectra. In \fig\ref{F8_realtime}, we show an example of real time data acquisition. In each of the four panels we show SN spectrum obtained using our DRS with an integration time of as low as 100 ms after the TTL trigger pulse. The magnetic field strength was also changed after each trigger pulse. Hence, we can sample the magnetic field with a time resolution of 100 msec. The integration time of the data shown in \fig \ref{F8_realtime} was 100 ms and the corresponding spectral resolution was $\approx$ 610 Hz resulting in an SNR
$\approx$ 5. Whereas for the data shown in \fig\ref{F7} the integration time was 1 s and the spectral resolution was 30 kHz, hence an SNR $\approx$ 15. 

For testing and verification of the triggered mode acquisition, we generated an external trigger TTL pulse from a function generator and fed this into a SMA-to-GPIO board, which was connected to the Red-Pitaya  using a ribbon cable and connector assembly. The acquisition time was set to $\approx$ 100 ms. The trigger rate was $\approx$ 10 per minute.

The output from the BPD was fed to the DRS. The triggered acquisition works as follows: (a) The FPGA waits for a rising edge on the trigger input port, (b) On the rising edge of the trigger input, a \texttt{dv} signal is asserted, upon which the BRAM starts to store the data, (c) Once 256 such samples are written, the data is packetized and transferred to the DAQ server. The trigger count and an absolute time stamp is included in the header data of the Ethernet packet for assisting in data analysis.

Another application of using our device is the temporal and spatial correlation measurements in a system with laser cooled atoms and ions. Intrinsically the correlation signals are expected to be extremely narrow in frequency domain and are promising candidates for various quantum technology applications. However, the measurement duration is typically limited to few milliseconds. Therefore, it is difficult to perform the SNS in cold atoms using traditional SFSA and our DRS emerges as a promising candidate for this purpose. In fact, we have already started using the DRS described in this article in cold atom measurements. Also, since in the RTDR configuration, the DRS records the voltage samples directly, it gives us the flexibility to achieve high frequency resolutions, which is only limited by the timing jitter of the on-board clock.

\begin{figure*}[!ht]
    \centering
    \includegraphics[trim={0.5cm 0cm 0.25cm 0.25cm},clip,width=\linewidth]{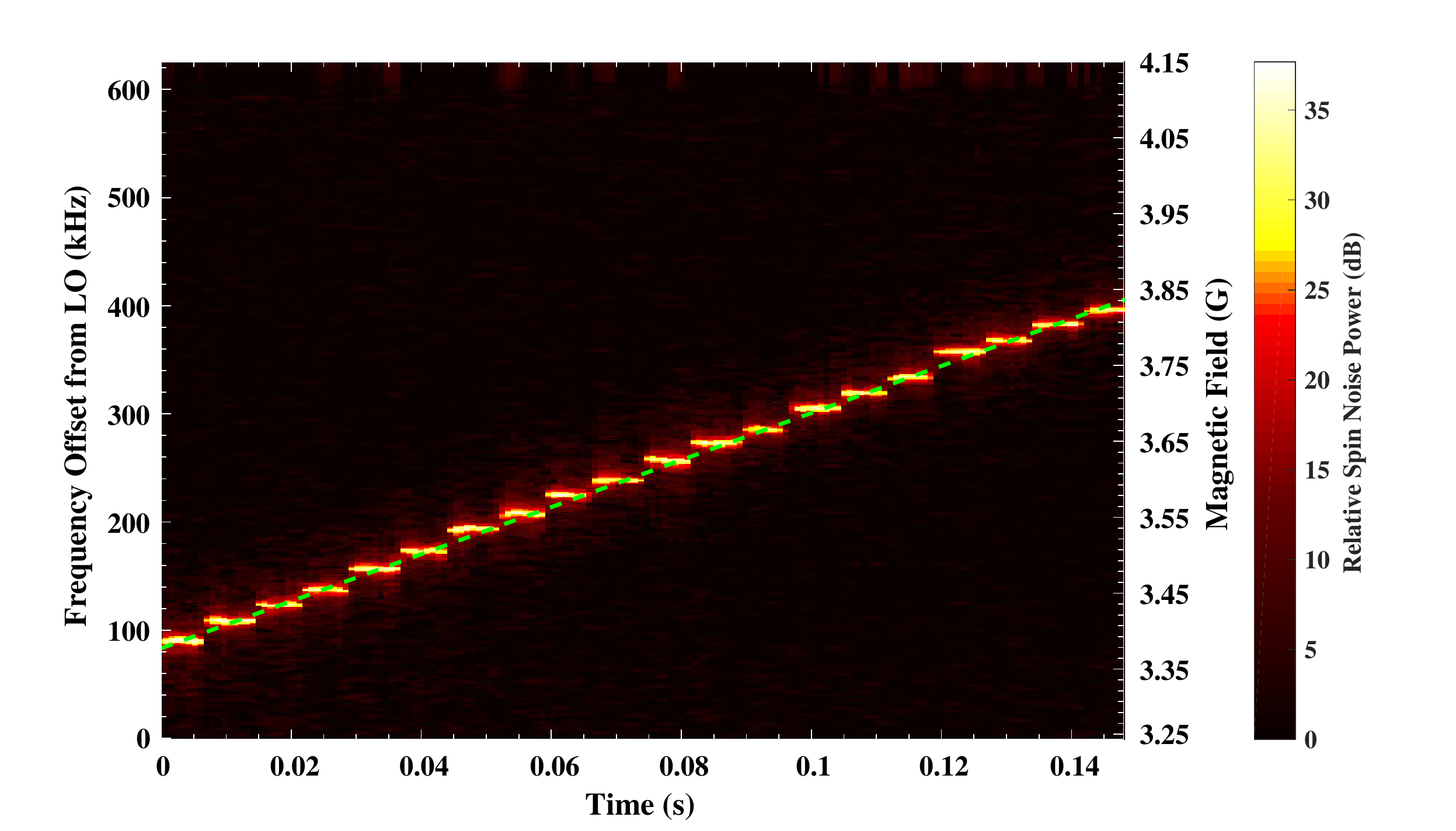}
    \caption{Waterfall plot of the DRS output when the magnetic field was swept from $\sim$ 3.35 G to 3.85 G. We have synchronously varied the coherent drive frequency (see text). The yellow patches represent the spin noise signal in the time-frequency plane, the green dashed line is a fit to the centroids of the yellow patches and its slope indicates the rate of variation of the magnetic field in time.}
    \label{F10}
\end{figure*}

\par
Our DRS in the RTDR mode, allows for a minimum time resolution of 800 ns when the data is treated in its raw form. When spectral analysis is carried out with a N point fourier transform, the channel width and the time resolution is $\frac{1.25 MHz}{N}$ and $800 ns \times N$, respectively. Therefore, when the signal is inherently strong, higher time resolutions can be obtained at the cost of reduced frequency resolution. However, for intrinsic SN signal from atomic vapor presented in this article, the SNR deteriorates for shorter integration times which affects the precision of the measurements. Therefore, for the purpose of demonstration, we integrate the SN signal for 100 ms time window and obtain a precision in measured magnetic field of the order of 800 $\mu$G.

To demonstrate the response of the detection system to fast varying magnetic fields, we conducted an experiment \cite{comment}, where we added a coherent drive \cite{fleischhauer2005} using a pair of Raman beams, which enhanced the SN signal strength million fold, improving the SNR. We then varied the magnetic field from 3.35 G to 3.85 G, synchronously with the coherent drive field frequency, and recorded the signal using our DRS. The results from this experiment is shown in Fig. 9, where each pixel along time and frequency axis is $\approx$ 800 $\mu s$ and 610 Hz, respectively. We see that the DRS simultaneously tracks the “step” changes in the magnetic field, as well as its drifts in milli-second timescales.

 We highlight that a time resolved measurement of magnetic field has applications ranging from geophysics \cite{prouty2013} to physiology\cite{bison2003,uchikawa1992}. The time stamp contained in the received data (see Section \ref{s3} B), can be used to determine the absolute time variation of the magnetic field strength. 

\section{Conclusions}\label{s6}
\par
We presented the development of a software defined digital receiver system (DRS) with two operating modes for spin noise spectroscopy (SNS) experiments. The highlight of this work is that we show the applicability of SNS in precision magnetometry and measure fast temporal variations in
the magnetic field. This receiver is  fully re-configurable and had a short development time at a low cost.
\par
The FFTS mode allows for user programmable integration times. The RTDR mode was specifically developed for high spectral and temporal resolution studies of spin noise (SN) signals from both hot and cold atoms. Such a mode, where high time resolution voltage data can be recorded, does not exist in SFSA. The FFTS can also be used to supplement the RTDR mode as follows: using the FFTS, the user will be able to explore the spectrum over a wide range of frequencies and once the peak location is known, the RTDR can be used to obtain a time-frequency resolved picture of the signal of interest. 
\par
Future directions for the system development include using a DDS core which can internally generate the LO signals to facilitate a two channel
implementation, replace the DAQ computer with a ARM based microprocessor system viz. Raspberry Pi for system miniaturization.
\par
The receiver described here was developed to be a part of a compact, portable SNS magnetometer. While we have demonstrated the application of the digital receiver using the SNS experiments, it can be easily adapted for use in other experiments requiring correlation measurements.

\section*{Acknowledgments}
The authors acknowledge  partial support provided by Ministry of Electronics and Information Technology (MeitY), Govt. of  India under grant for ``Center for Excellence in Quantum Technology"  with Ref. No. 4(7)/2020-ITEA. Authors also acknowledge   Department of Science and Technology (DST), Govt. of India,  Prof. Hema Ramachandran, Prof. Dibyendu Roy,  Electronics Engineering Group (EEG) and central workshop facility, Raman Research Institute. We extend our thanks to the CASPER signal processing community and Dr. Pavel Demin for maintaining many useful open-source IP cores. We are grateful to Xilinx for donation of Vivado Design Suite, through their University Program.

\bibliographystyle{IEEEtran}
\bibliography{IEEEbibliography}

\begin{thebibliography}{10}
\bibitem{comment}
``The details of the experiment and its applicability in precision magnetometry
  will be described in a subsequent publication.''
\end{thebibliography}

\end{document}